\documentclass[aps,prd,twocolumn,showpacs,groupedaddress,nofootinbib,amsmath]{revtex4}
\usepackage{graphicx}
\usepackage{epsfig}
\usepackage{multirow}

\begin{document}

\title{Nature of Roberge-Weiss transition end points for heavy quarks in $N_f=2$ lattice QCD with  Wilson fermions }

\author{Liang-Kai Wu}
\thanks{Corresponding author. Email address: wuliangkai@163.com}
\affiliation{Faculty of Science, Jiangsu University, Zhenjiang 212013, People¡¯s Republic of China}
%\affiliation{National supercomputer center, Tianjin, 300457, People¡¯s Republic of China}

\author{Xiang-Fei Meng}
\affiliation{National Supercomputer Center, Tianjin, 300457, People¡¯s Republic of China}

\date{\today}

\begin{abstract}
The phase structure of QCD with imaginary chemical potential provides information on the phase diagram of QCD  with real chemical potential. With imaginary chemical potential $i\mu_I=i\pi T$,
previous studies show that the Roberge-Weiss (RW) transition end points  are triple points at both large and small quark masses,  and second order transition points at intermediate quark masses. The triple and second order end points are separated by two tricritical ones.  We present simulations with $ N_f=2 $  Wilson fermions to investigate the nature of  RW  transition end points.
The simulations are carried  out at 8 values of  the hopping parameter $\kappa$ ranging from 0.020 to 0.140 on different lattice volumes. The Binder cumulant,  susceptibility and reweighted distribution
of the imaginary part of Polyakov loop are employed to determine the nature of RW transition end points. The simulations show that  the two tricritical points are within the range $0.070-0.080$ and $0.120-0.140$, respectively.
\end{abstract}


\maketitle

\section{INTRODUCTION}
\label{SectionIntro}

The study  of QCD phase diagram is of great importance theoretically and phenomenologically, overviews may be found in Ref.~\cite{Fukushima:2010bq,Fukushima:2011jc} and references therein. On and below the
scales of a baryon mass  which is relevant to heavy ion collision and astrophysics,
 the non-perturbative nature of QCD   warrants the Monte Carlo simulation of lattice QCD.
 Although substantial progress has been achieved  at zero baryon density,  the MC simulation of lattice QCD is accompanied by the "sign" problem when studies are extended to finite density, for example, see Ref.~\cite{Kogut:2007mz}.
  To date
many indirect methods have been proposed to circumvent the "sign" problem, overviews with
references to these methods can be found in Ref.~\cite{Kogut:2007mz,Schmidt:2006us}. One of these methods consists of simulating
QCD with  imaginary chemical potential for which the fermion determinant is
positive~\cite{deForcrand:2010he,D'Elia:2009tm,D'Elia:2009qz,Nagata:2011yf,Bonati:2010gi,D'Elia:2007ke,Wu:2006su,deForcrand:2006pv,deForcrand:2008vr}.

QCD with imaginary chemical potential has a rich phase structure, and it not only deserves detailed investigations in its own
right theoretically, but also has significant relevance to physics at zero or small real chemical potential\cite{D'Elia:2009qz,D'Elia:2007ke,Bonati:2010gi,Kouno:2009bm,Sakai:2009vb,deForcrand:2010he,Aarts:2010ky,Philipsen:2010rq,Bonati:2012pe}.
The tricritital
 line found at the imaginary chemical potential, with its associated scaling law, imposes constraints on the phase diagram of QCD at real chemical potential~\cite{deForcrand:2010he}.

The partition function of QCD with complex chemical potential has two important symmetries~\cite{Roberge:1986mm}: reflection symmetry in $\mu=\mu_R+i\mu_I$ and
periodicity in imaginary chemical potential.
 The $Z(3) $ symmetry  is explicitly broken at the presence of dynamical quarks for real chemical potential. However, for complex $\mu$,  due to the
 periodicity of partition function in imaginary chemical potential,
the $Z(3) $ symmetry is restored.  Different Z(3) sectors are distinguished  by the Polyakov loop. Transition between adjacent Z(3) sectors in $\mu_I$
is analytic  for low temperature, and is of  first order (RW transition) for high temperature. The
first order transition  that takes place at critical values
of the imaginary chemical potential $\mu_I/T =
(2n+1){\pi}/3$~\cite{Roberge:1986mm,deForcrand:2002ci,Lombardo} forms a transition line,
thus  the first order transition line in the high temperature region necessarily ends at an end point  $T_{RW}$ when the temperature is decreased sufficiently low.

Recent numerical studies~\cite{deForcrand:2010he,D'Elia:2009qz,Bonati:2010gi} show that the RW transition end points
are  triple points for small and heavy quark masses,
and second order points for intermediate quark masses. So there exist two tricritical points  separating the first order
transition points from the second ones. Moreover, it is pointed out~\cite{deForcrand:2010he,Philipsen:2010rq,Bonati:2012pe} that the scaling behaviour at the tricritical points
may shape the the critical line which separate different transition region for real chemical potential, and thus, the critical line for real chemical potential is expected to be
qualitatively consistent with the scenario suggested in Ref.~\cite{deForcrand:2006pv,deForcrand:2008vr} which show that the first order transition region shrinks with increasing real chemical potential.

Most of studies for finite density QCD have been performed using
staggered fermion action or the improved
versions.
%~\cite{Karsch,Cheng:2006qk,Cheng:2006aj,Bernard:2004je,Aoki:2006br,Bernard:2007nm,Aoki:2005vt},
 The disadvantages of staggered fermion discretization~\cite{Heller:2006ub,Bunk,Golterman} warrants studies of lattice QCD with a  different
 discretization. In Ref.~\cite{Philipsen:2014rpa,Wu:2013bfa}, Wilson fermion  is employed to investigate
 the nature of the  RW transition end points, but the $\kappa$ values in Ref.~\cite{Wu:2013bfa} are limited from  $0.155$ to $0.198$.

In this paper, Proceeding  with our previous work~\cite{Wu:2013bfa} along this direction,  we attempt to investigate the RW transition line end points  with $N_f=2$ Wilson fermions with $\kappa$  ranging from $0.020$ to $0.140$. In Sec.~\ref{SectionLattice},
we define the lattice action with imaginary chemical potential and
the physical observables we calculate.  Our simulation results are
presented in Sec.~\ref{SectionMC} followed by discussions in
Sec.~\ref{SectionDiscussion}.

\begin{figure*}[t!]
\includegraphics*[width=0.49\textwidth]{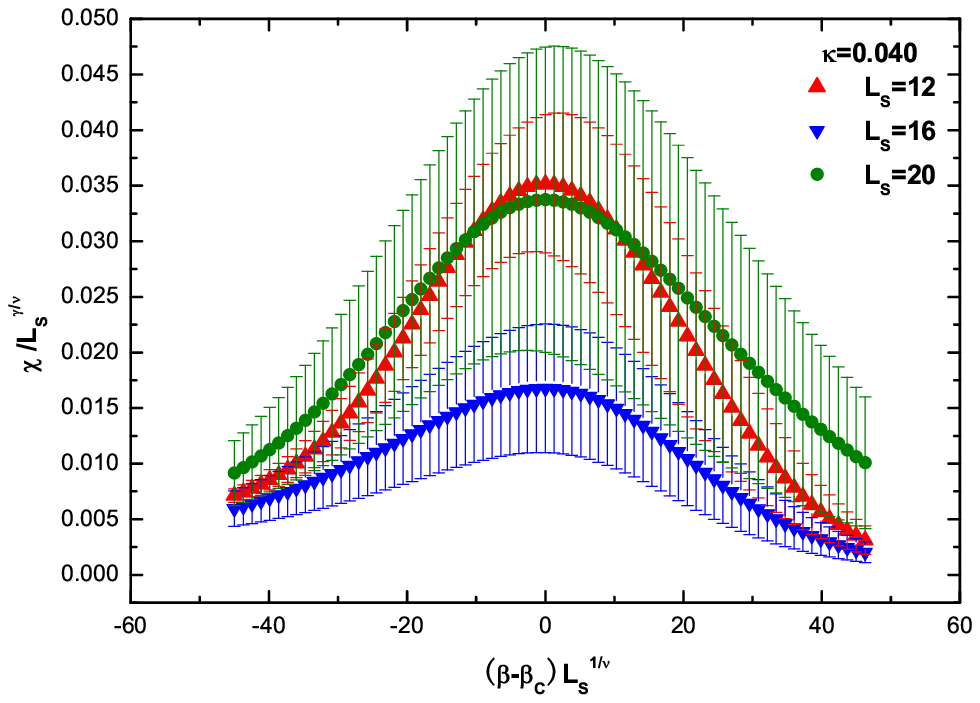}
\includegraphics*[width=0.49\textwidth]{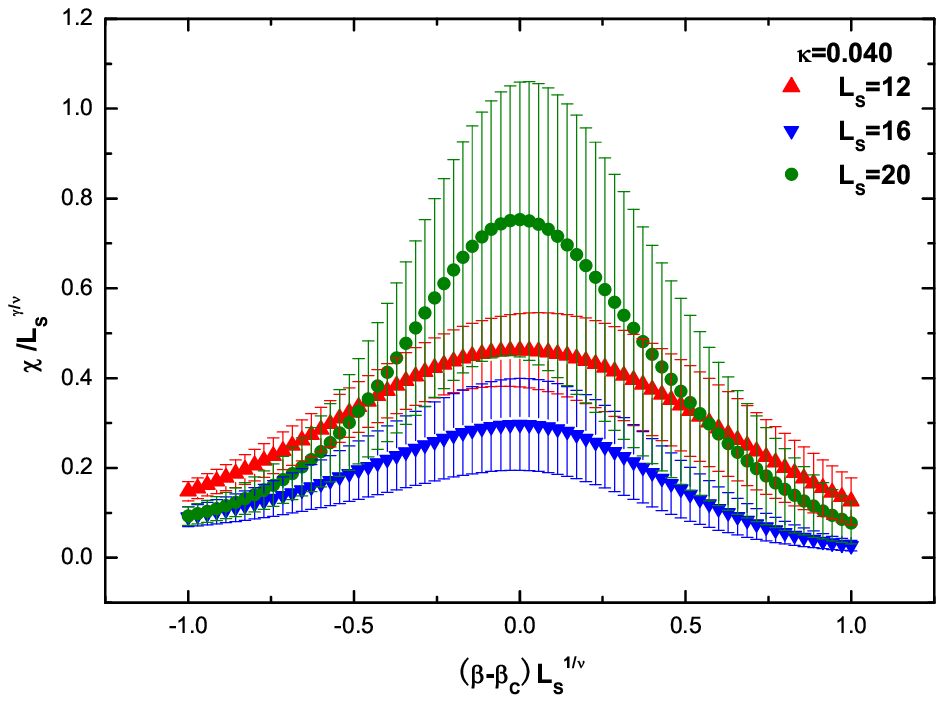}\\
\caption{\label{fig1} Scaling behavior of the susceptibility of  imaginary part of  Polyakov loop
according to  first order critical index (left panel), and to  3D Ising critical index (right panel) at $\kappa=0.040$.}
%\vspace*{-2cm}
\end{figure*}

\begin{figure*}[t!]
\includegraphics*[width=0.49\textwidth]{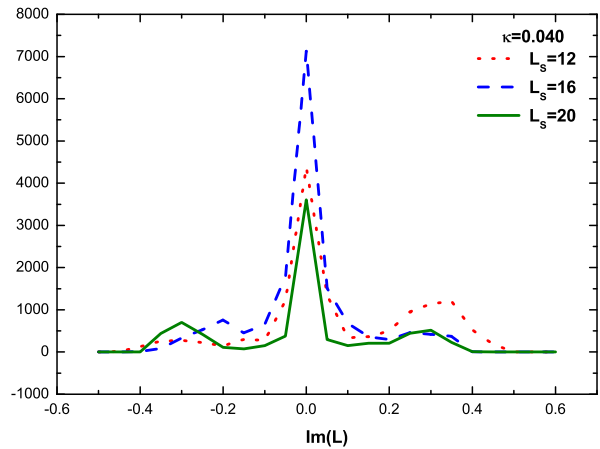}
\caption{Reweighted distribution of the imaginary part
of Polyakov loop at $\kappa=0.040$  at the corresponding end point $\beta_{RW}$.}
\label{fig2}
\end{figure*}

\begin{figure*}[t!]
\includegraphics*[width=0.49\textwidth]{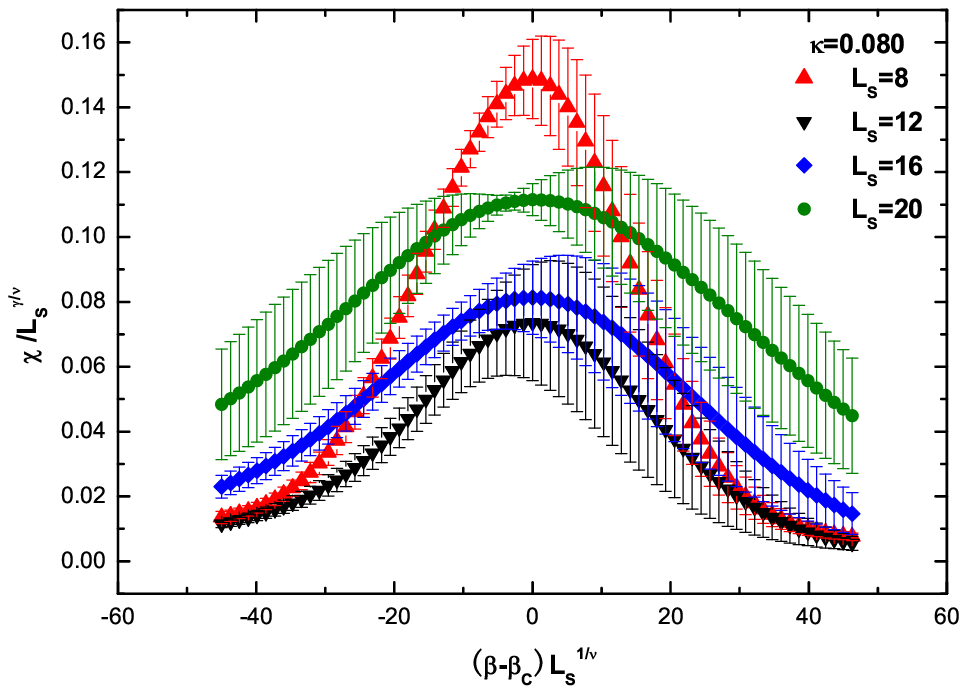}
\includegraphics*[width=0.49\textwidth]{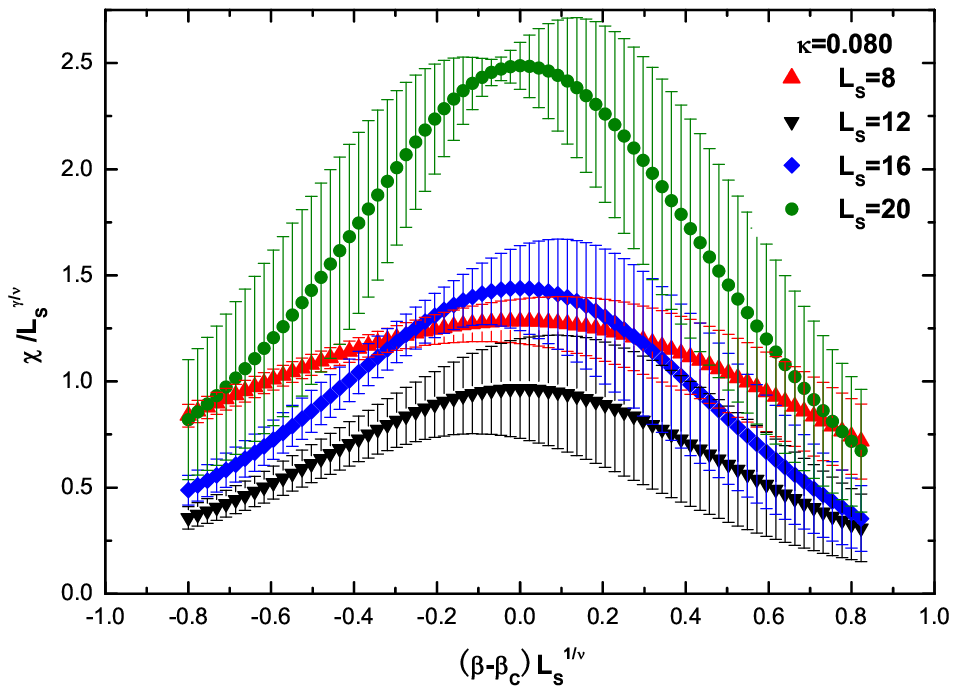}
\caption{\label{fig3} Scaling behavior of the susceptibility of imaginary part of  Polyakov loop
according to  first order critical index (left panel), and to  3D Ising critical index (right panel)at $\kappa=0.080$.}
%\vspace*{-2cm}
\end{figure*}

\begin{figure*}[t!]
\includegraphics*[width=0.49\textwidth]{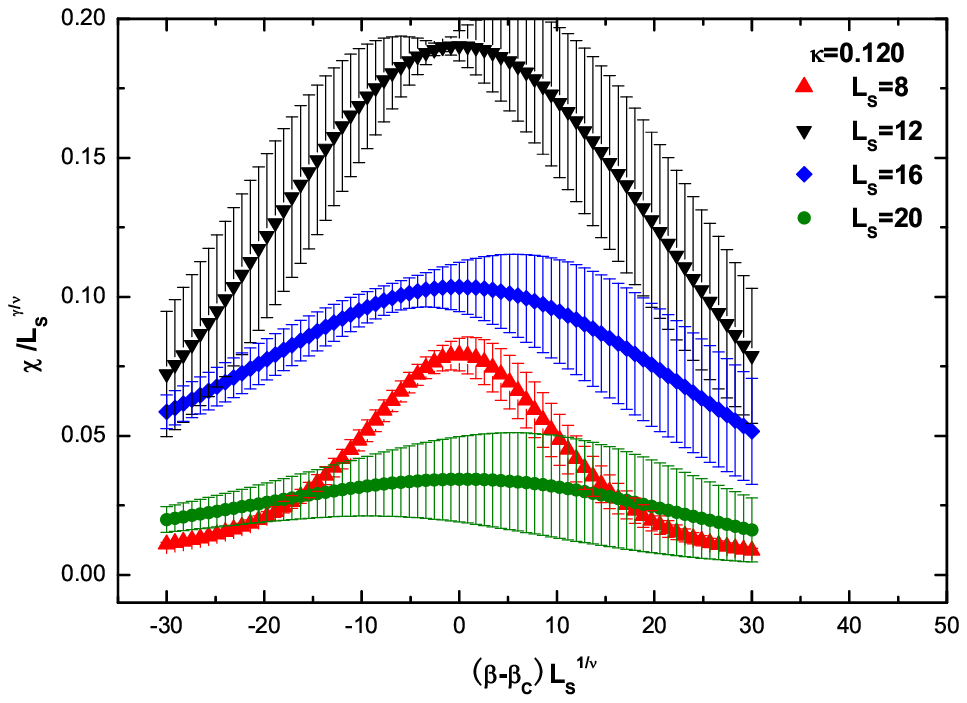}
\includegraphics*[width=0.49\textwidth]{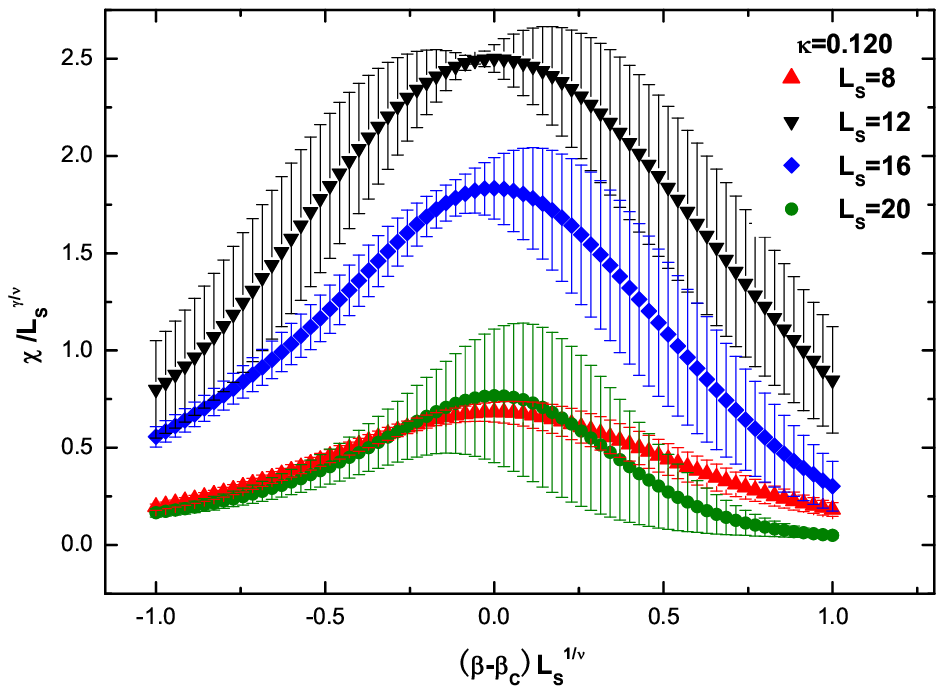}\\
\caption{\label{fig4} Scaling behavior of the susceptibility of  imaginary part of  Polyakov loop
according to  first order critical index (left panel), and to  3D Ising critical index (right panel)at $\kappa=0.120$.}
%\vspace*{-2cm}
\end{figure*}

\begin{figure*}[t!]
\includegraphics*[width=0.49\textwidth]{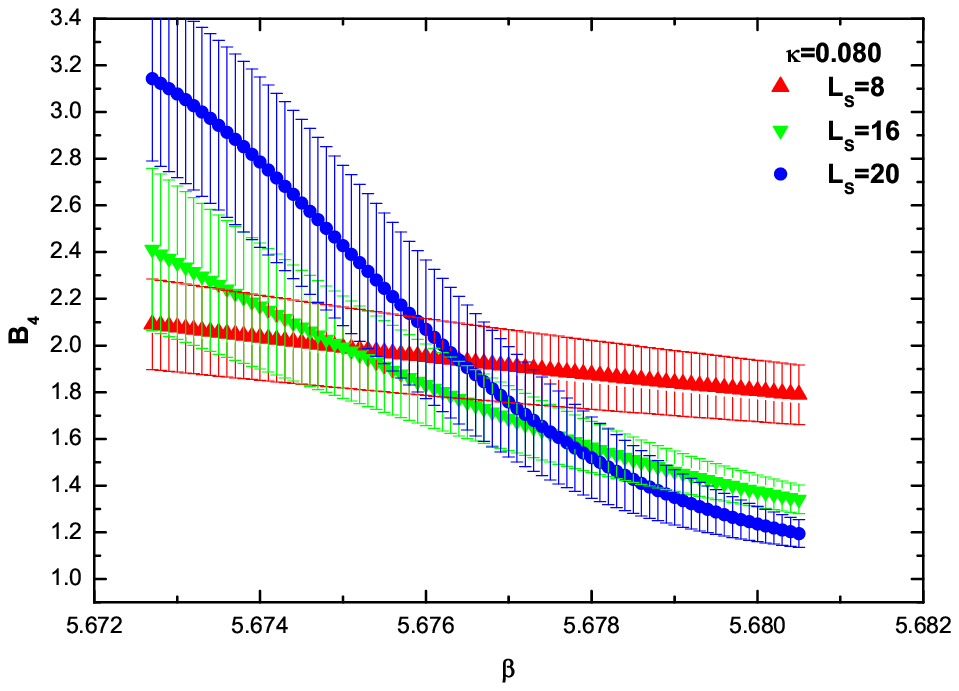}
\includegraphics*[width=0.49\textwidth]{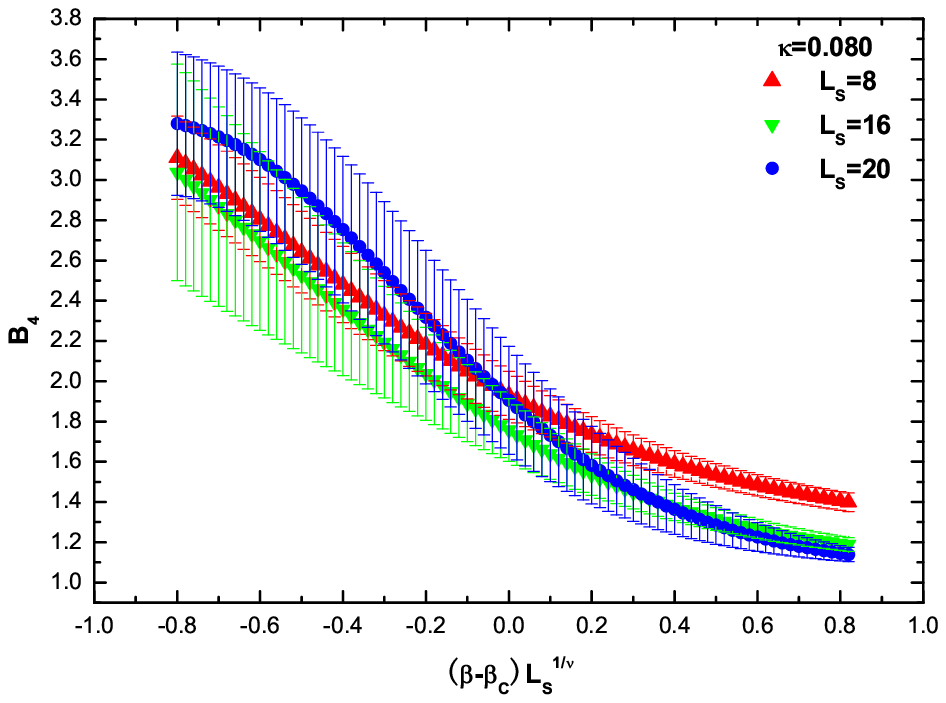}\\
\includegraphics*[width=0.49\textwidth]{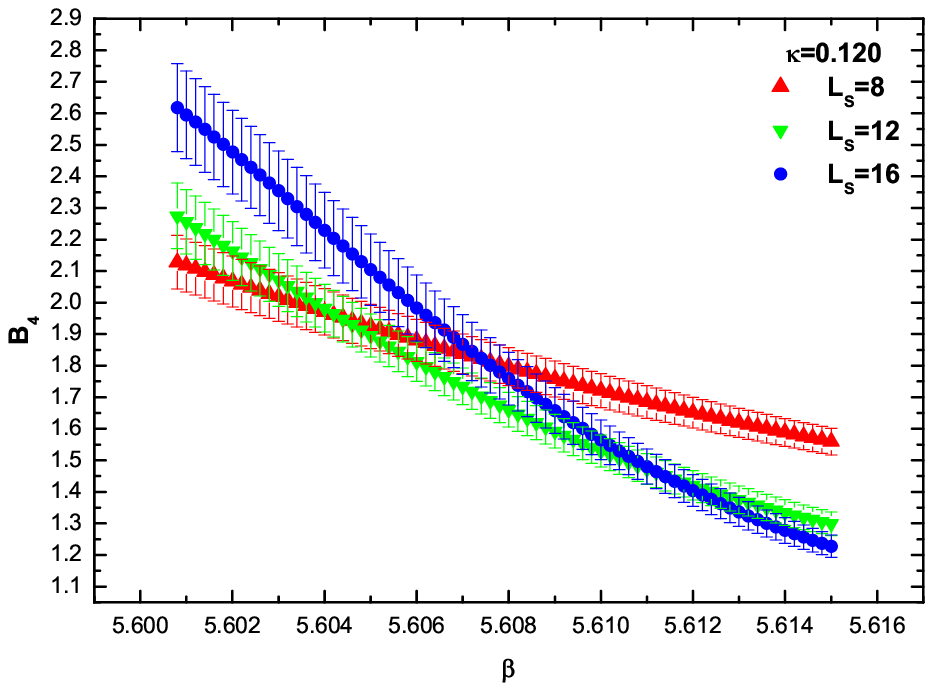}
\includegraphics*[width=0.49\textwidth]{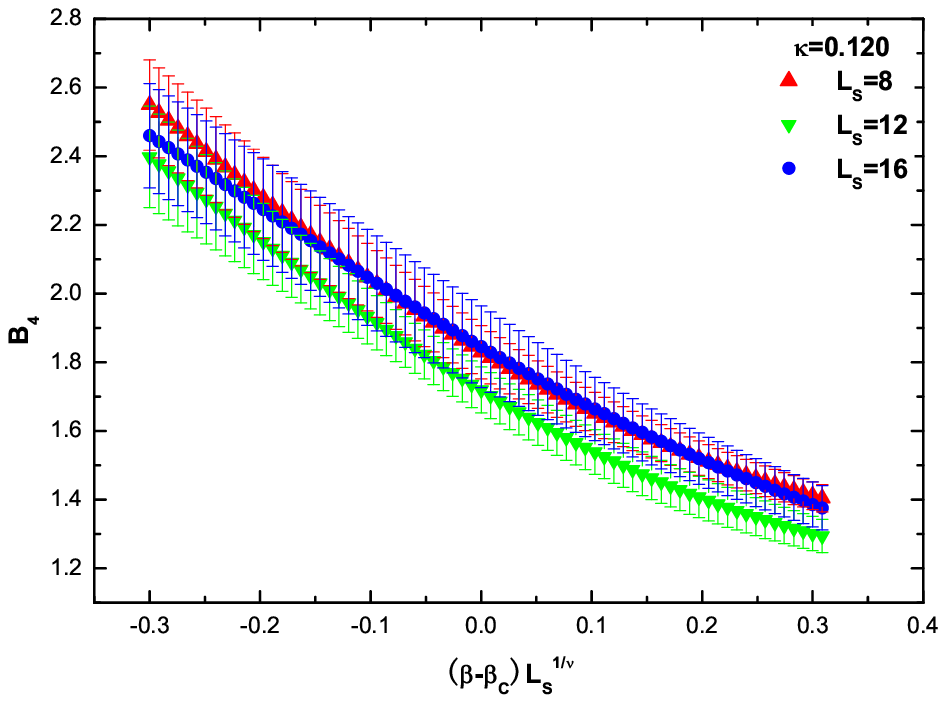}
\caption{Binder cumulants as  a function of $\beta$ on various spatial volume intersect at one point (left panels),
and as  a function of $(\beta-\beta_c)L_s^{1/\nu}$ with values of $\beta_c$, $\nu$ from Table.~\ref{critical_beta_B4} collapse (right panels).}
\label{fig5}
\end{figure*}

\begin{figure*}[t!]
\includegraphics*[width=0.49\textwidth]{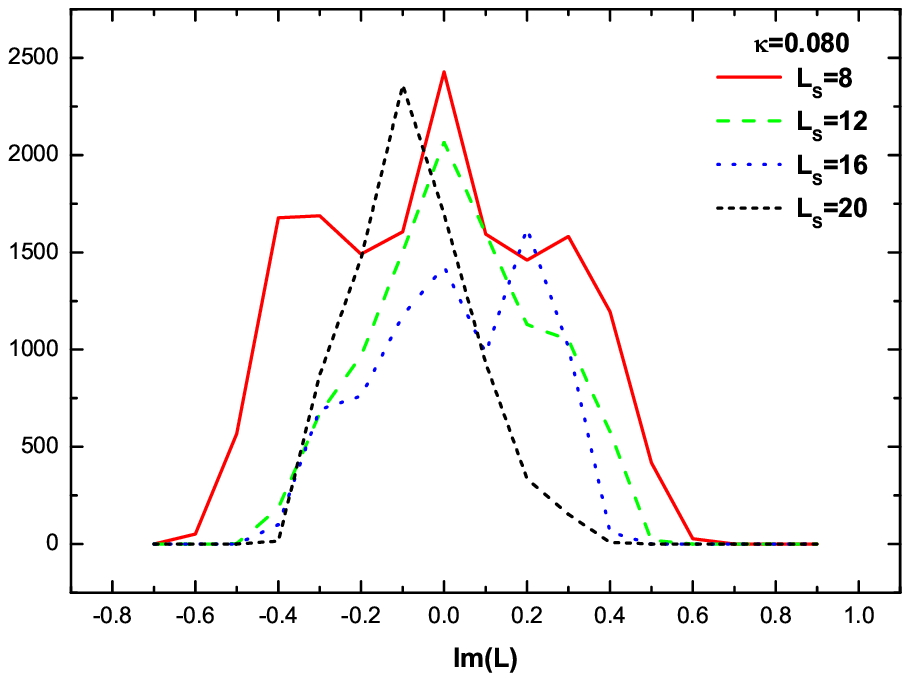}
\includegraphics*[width=0.49\textwidth]{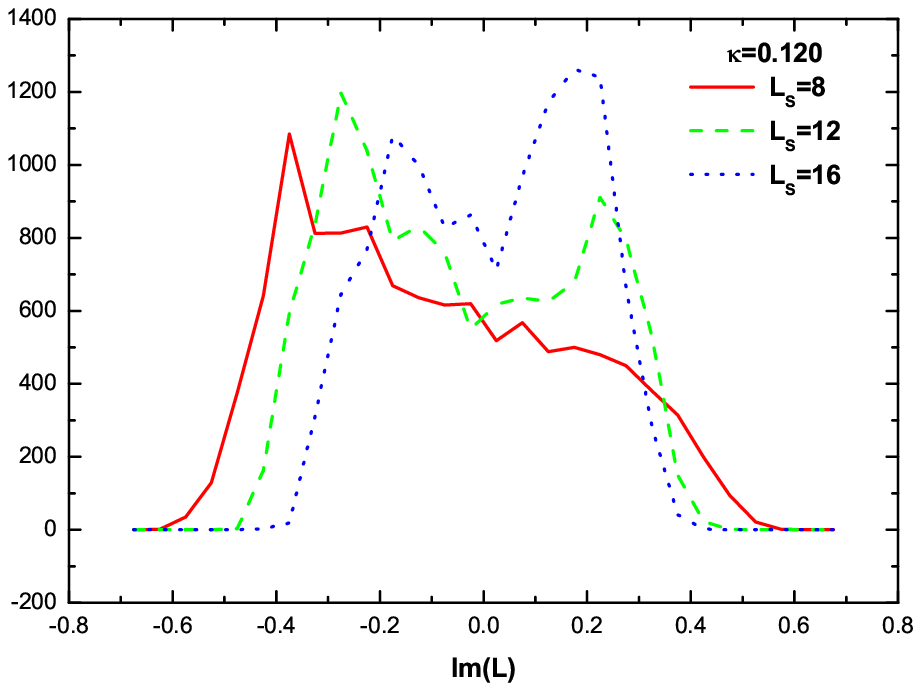}
\caption{\label{fig6} Reweighted distribution of the imaginary part
of  Polyakov loop at $\kappa=0.080,\,0.120$ at the corresponding end points $\beta_c$ which are extracted by fitting.}
%\vspace*{-2cm}
\end{figure*}

\begin{figure*}[t!]
\includegraphics*[width=0.49\textwidth]{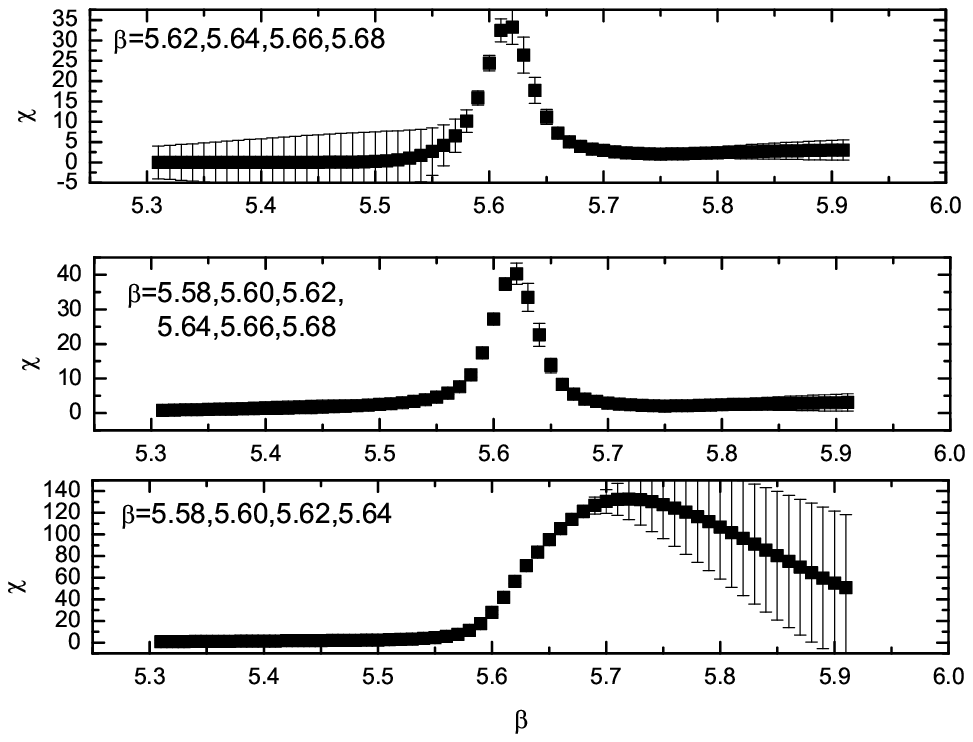}
\includegraphics*[width=0.49\textwidth]{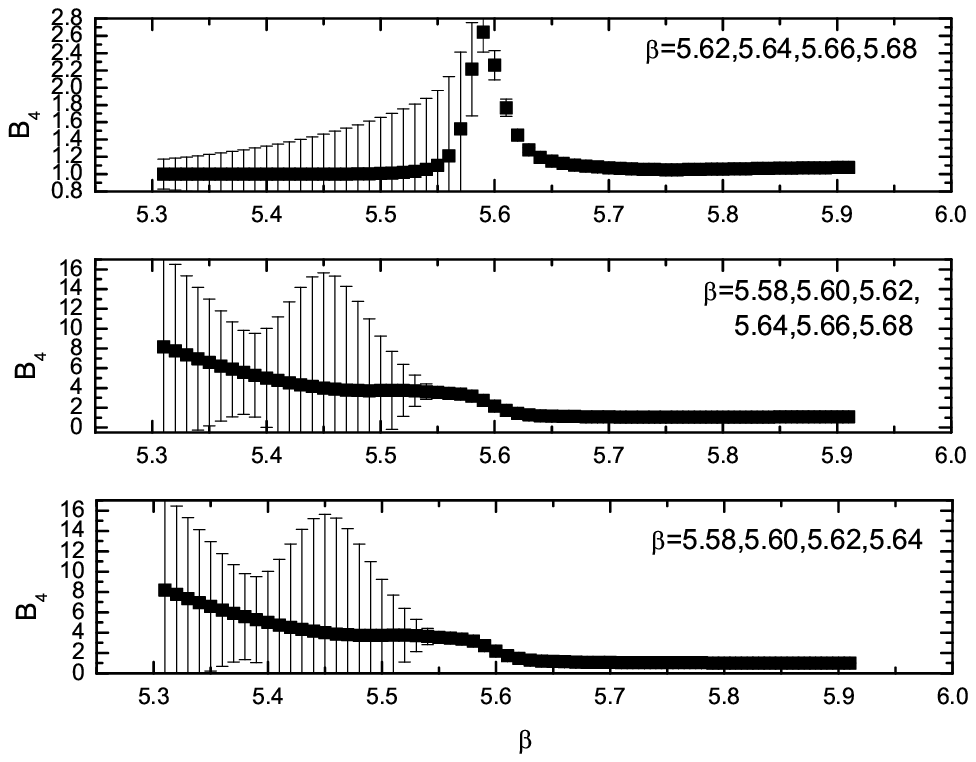}
\caption{\label{fig7} Behavior of susceptibility of  imaginary part of  Polyakov loop (left panels) and Binder cumulant (right panels) with different selection of $\beta$ values on $L_s=8$ at $\kappa=0.120$. }
%\vspace*{-2cm}
\end{figure*}

\section{LATTICE FORMULATION WITH IMAGINARY CHEMICAL POTENTIAL}
\label{SectionLattice} We consider a system with
$N_f=2$ degenerate Wilson fermions whose  partition function   with chemical potential is
\begin{eqnarray}
\label{QCD_partition}
Z &= &\int [dU][d\bar\psi][d\psi]e^{-S_g-S_f} \nonumber \\
&=& \int [dU] \biggl({\rm Det} M [U,\theta]\biggr)^{N_f} e^{-S_g}.
\end{eqnarray}
where $S_g$ is the gauge action, and $S_f$ is the quark
action with the quark imaginary chemical potential $\mu_I=\theta T$.  For  $S_g$, we use the
standard one-plaquette action
\begin{eqnarray}
S_g=\beta\sum_p \biggl(1-\frac{1}{N_c}{\rm ReTr}U_p\biggr),
\end{eqnarray}
where $\beta=6/g^2$, and the plaquette variable $U_p$ is the ordered
product of link variables $U$ around an elementary plaquette. For
$S_f$, we use the the standard Wilson action
\begin{eqnarray}
S_f=\sum_{f=1}^{N_f}\sum_{x,y} {\bar
\psi}_f(x)M_{x,y}(U,\kappa,\mu){\psi}_f(y),
\end{eqnarray}
where $\kappa$ is the hopping parameter, related to the bare quark
mass $m$ and lattice spacing $a$ by $\kappa=1/(2am +8)$. The fermion
matrix is
\begin{eqnarray}
M_{x,y}(U,\kappa,\mu) & &  = {\delta_{x,y}} - \kappa \sum_{j=1}^{3}
\bigg[
(1-\gamma_{j})U_{j}(x)\delta_{x,y-\hat{j}} \nonumber \\
& &+ (1+\gamma_{j})U_{j}^{\dagger}(x-\hat{j})\delta_{x,y+\hat{j}}
\bigg]
\nonumber \\
&&- \kappa
   \bigg[(1-\gamma_{4})e^{a\mu}U_{4}(x)\delta_{x,y-\hat{4}}
\nonumber \\
&&+
   (1+\gamma_{4})e^{-a\mu}U_{4}^{\dagger}(x-\hat{4})\delta_{x,y+\hat{4}}\bigg].
\end{eqnarray}

We carry out simulations at $\theta=\pi$. As it is pointed out that the system is invariant under the charge
 conjugation at $\theta =0,\pi$, when $\theta $ is fixed~\cite{Kouno:2009bm}. But the $\theta$-odd quantity $O(\theta)$ is not invariant at $\theta=\pi$
 under charge conjugation. When $T < T_{RW}$, $O(\theta)$ is a smooth function of $\theta$, so it is zero at $\theta=\pi$.
 Whereas when $T > T_{RW}$,  the two charge violating solutions cross each other at $\theta=\pi$. Thus the charge symmetry is
 spontaneously broken there and the $\theta$-odd quantity $O(\theta)$ can be taken as order parameter . In this paper, we take the imaginary part of Polyakov loop as the order parameter.

The Polyakov loop $ L $ is defined as the following:
\begin{eqnarray}
\langle L \rangle=\left\langle \frac{1}{V}\sum_{\bf x}{\rm  Tr}
\left[ \prod_{t=1}^{N_t} U_4({\bf x},t) \right] \right\rangle ,
\end{eqnarray}
here and in the following, $V$ is the spatial lattice volume. To simplify the notations, we use $X$ to represent
the imaginary part of Polyakov loop $L$, $X={\rm Im}(L)$.

The susceptibility of imaginary part of Polyakov loop $\chi$ is defined as
\begin{eqnarray}
\chi= V \left\langle( X - \langle  X\rangle)^2\right\rangle ,
\end{eqnarray}
which is expected to scale as:~\cite{D'Elia:2009qz,Bonati:2010gi}
\begin{eqnarray}\label{chi_scaling}
\chi= L_s^{\gamma/\nu}\phi(\tau L_s^{1/\nu}) ,
\end{eqnarray}
where $\tau $ is the reduced temperature $\tau=(T-T_{RW})/T_{RW}$, $V=L_s^3$ .
This means that the curves $\chi/L_s^{\gamma/\nu}$ at different lattice volume should collapse with the same curve when plotted against $\tau L_s ^{1/\nu}$.
In the following, we employ $\beta-\beta_{RW}$ in place of $\tau=(T-T_{RW})/T_{RW}$. The critical exponents relevant to our study
are collected in Table.~\ref{critical_exponents}~\cite{Bonati:2010gi,Pelissetto:2000ek}.
%\begin{widetext}
\begin{table}[htmp]
%\begin{ruledtabular}
\begin{center}
\begin{tabular}{|c|ccc|}
\hline
          &  $\nu$     & $\gamma $    &       $\gamma/\nu$    \\
 \hline
3D ising  & 0.6301(4)  & 1.2372(5)    &        1.963                  \\
tricritical & 1/2      &  1           &           2                         \\
 first order    & 1/3  &  1           &               3                        \\
 \hline
\end{tabular}
\end{center}
%\end{ruledtabular}
\caption{\label{critical_exponents}Critical exponents relevant to our study.}
\end{table}

We also consider the Binder cumulant of  imaginary part of Polyakov loop which is defined as the following:
\begin{eqnarray}\label{binder_scaling}
B_4=\left\langle ( X - \langle  X\rangle)^4\right\rangle /
    \left\langle ( X - \langle  X\rangle)^2\right\rangle^2 ,
\end{eqnarray}
with  $\langle X \rangle =0$.
In the vicinity of the RW transition line
end points, $B_4$ with the finite size correction is a function of $x=(\beta-\beta_{RW})L_s^{1/\nu}$ and can be expanded as a series~\cite{deForcrand:2010he,Philipsen:2010rq,Bonati:2012pe}.
\begin{eqnarray}\label{binder_scaling_02}
B_4=B_4(\beta_c,\infty)+a_1x+a_2x^2+\cdots.
\end{eqnarray}
 In the thermodynamic limit, the critical index $\nu$ takes on the corresponding value summarized in Table.~\ref{critical_exponents}.
 $B_4(\beta_c,\infty)$
takes on the values 3,  1.5, 1.604, 2 for crossover, first order triple point, 3D Ising and tricritical transitions, respectively.
However, on finite spatial volumes, the steps of $B_4(\beta_c,\infty)$ are smeared out to continuous functions.

\section{MC SIMULATION RESULTS}
\label{SectionMC}

In this section, we first present the MC simulation parameters.  The $\phi$ algorithm with a Metropolis accept/reject
step  is used~\cite{Gottlieb:PRD:35:3972}.    The simulations are performed  at $\kappa=0.020,\,$ $0.040,\,$ $0.060,\,$ $0.070,\,$ $0.080,\,$ $0.100,\,$ $0.120,\,$ $0.140 $ on $L_t=4$ lattice.
 For each $\kappa$ value,  Lattices with  spatial size $L_s=8,\, 12,\, 16, \, 20$ are employed except
 at $\kappa=0.140$ where simulations are carried out on lattice size $L_s=8,\, 12$.

 The acceptance rates range within $53\%-95\% $.  $20-50$ molecular steps  are taken for each trajectory.
 $20,000-100,000$ trajectories are generated after $10,000-50,000$  trajectories as warmup.

The conjugate gradient method  is used to evaluate the fermion matrix
inversion. For each lattice size, we make simulations at  4-11 different $\beta$ values. In order to fill in  observables   at additional $\beta$ values,
we employ the Ferrenberg-Swendsen reweighting
method~\cite{Ferrenberg:1989ui}.

The critical coupling $\beta_{RW}$'s on various spatial volume at different $\kappa$ are summarized in Table.~\ref{critical_beta}. These $\beta_{RW}$'s are determined from the locations of  peak susceptibility of imaginary part of Polyakov loop.
\begin{table}[htmp]
\caption{\label{critical_beta}Results of critical couplings $\beta_{RW}$ on different spatial volume at different $\kappa$. }
\begin{ruledtabular}
\begin{center}
\begin{tabular}{c|cccc}

$\kappa$    &     $8 $        &         $  12   $       &    $16 $       & $  20 $       \\  \hline
0.020       &   $5.695(20)$   &          $5.697(14)$    & $5.691(7)$     & $5.684(10)$  \\
0.040       & $5.706(19) $    &         $ 5.694(12) $   & $5.688(9) $    &  $5.693(6) $             \\
0.060       & $5.725(25)$     &           $ 5.691(8)$   &  $5.687(7)$    & $5.689(6)$  \\
0.070       & $5.712(21) $    &            $5.690(9)$   & $ 5.695(5)$    & $5.693(6)$  \\
0.080       & $5.713(25)$     &           $5.687(13)$   & $ 5.683(7)$    & $5.684(5)$   \\
0.100       & $5.672(14)$     &            $5.659(6)$   & $ 5.687(4)$    & $5.688(3)$   \\
0.120       & $5.618(12)$     &            $5.650(7)$   & $ 5.619(3)$    &   $5.609(5)$            \\
0.140       & $5.639(21)$     &           $5.636(21)$   & $ -$    & $-$              \\
\end{tabular}
\end{center}
\end{ruledtabular}
\end{table}

%%%%%%%%%%%%%%%%%%%%%%%%%%%%%%%%%%%%%%%%%%%%%%%%%%%%%%%%%%%%%%%%%%%%%%%%%%%%%%%%%%%%%%%%%%
%%%%%%%%%%%%%%%%%%%%%%%%%%%%%%%%%%%%%%%%%%%%%%%%%%%%%%%%%%%%%%%%%%%%%%%%%%%%%%%%%%%%%%%%%%

\begin{table*}[htmp]
\caption{\label{critical_beta_B4}Results of critical couplings $\beta_{RW}$ and the critical index $\nu$ by fitting Eq.~(\ref{binder_scaling_02}) to data on different spatial volume. The errors of $\beta_{RW}$ are very small, so we take them to be zero.   }
\begin{ruledtabular}
\begin{center}
\begin{tabular}{c|ccccccc}

$\kappa$    &     $L_s $      &   $  \beta_{RW}$  & $\nu $         & $  B_4(\beta_c,\infty) $  & $a_1$  &  $a_2$  & r-square    \\  \hline
0.060       &$8 \,,12 \,,16$        &   $ 5.6838(0)$    & $0.3374(3)$    & $ 2.3189(4)$   &  $-0.0793(5)$ & $0.00127(1)$  &   0.990  \\
0.070       &$8 \,,12\,, 16$        &   $ 5.6812(0)$    & $0.4187(2)$    & $ 2.0745(3)$   &  $-0.289(8)$  & $0.0244(2)$    &   0.963  \\
0.080       &$8\,, 16\,, 20$        &   $ 5.6753(0)$    & $0.5872(4)$    & $ 1.9747(3)$   &  $-1.272(4)$ & $0.427(3)$     &   0.977  \\
0.100       &$8\,, 12\,, 16$        &   $ 5.6524(0)$    & $0.5218(8)$    & $ 1.8405(6)$   &  $-0.6956(56)$ & $0.185(3)$     &   0.995  \\
0.120       &$8\,, 12\,, 16$        &   $ 5.6072(0)$    & $ 0.6791(4)$   & $1.7964(8)$    &  $-1.880(25)$ & $1.364(36)$ &   0.980  \\
0.140       &$8\,, 12\     $        &   $ 5.5428(21)$   & $ 0.2222(2)$   & $1.1853(4)$    &  $-0.0005(3)$ & $0(0)$ &   0.893  \\
\end{tabular}
\end{center}
\end{ruledtabular}
\end{table*}

 We present the rescaling susceptibility of imaginary part of Polyakov loop $\chi/{{L_s}^{\gamma/\nu}}$ as a function of $(\beta-\beta_{RW})L_s^{1/\nu}$ in Fig.~\ref{fig1}. At $\kappa=0.020,\,0.040$, the fermion mass is very large, the deconfinement transition behaviour is expected to be mainly governed by the
first order transition of pure gauge system. However, from  Fig.~\ref{fig1}, we can find that
$\chi/{{L_s}^{\gamma/\nu}}$ according to the first order transition index  or 3D Ising transition index at $\kappa=0.040$  does not collapse with the same curve.
 The reweighted distribution of imaginary part of Polyakov loop presented in
Fig.~\ref{fig2} which exhibits one-state signal other than weak two-state signal is not favor of first order transition.   We can not determine the nature of transition  decisively based on  the simulation results at $\kappa=0.040$.  At $\kappa=0.020$, similar behaviour is observed.

 The rescaling behavior of $\chi/{{L_s}^{\gamma/\nu}}$  at $\kappa=0.080$ and $\kappa=0.120$ is presented in Fig.~\ref{fig3} and Fig.~\ref{fig4}, respectively. From the two  panels of
Fig.~\ref{fig3}, we can find that neither the first order transition index nor 3D Ising transition index is suitable to describe  the system at RW transition  end point. At $\kappa=0.060, \,$ $ 0.070,\,$ $ 0.100,\, $ $ 0.120 $ the rescaling observable  $\chi/{{L_s}^{\gamma/\nu}}$  exhibit the similar behaviour as  that at $\kappa=0.080$.  For clarity, we  only present the results at $\kappa=0.120$ in Fig.~\ref{fig4}.

 In order to discern the scaling behaviour at $\kappa=0.060,\,$ $0.070,\,$ $0.080,\,$ $0.100,\,$ $0.120,\,$ $0.140$,
we turn to investigate Binder cumulant $B_4$ as defined in Eq.~(\ref{binder_scaling}) whose scaling behaviour is described in Eq.~(\ref{binder_scaling_02}). $B_4$ decreases with the increase of   $\beta$, and at one fixed $\kappa$ value, $B_4$ as a function of $\beta$ on various spatial volume is expected to intersect at one point. The intersection gives an estimate of accurate location of $\beta_{RW}$. By  fitting  to
Eq.~(\ref{binder_scaling_02}), we can extract critical index $\nu, \, \beta_{RW}$ and $B_4(\beta_c,\infty)$. The results are collected in Table.~\ref{critical_beta_B4}. We present $B_4$ as a function of $\beta$  at $\kappa=0.080,\,$ $0.120 $ in the left panels of Fig.~\ref{fig5},
and  $B_4$ as a function of $(\beta-\beta_{RW})L_s^{1/\nu}$ in the right panels of Fig.~\ref{fig5} with $\nu$ taken to be the extracted value through  fitting procedure
.
  From Table.~\ref{critical_beta_B4},  we find that the critical index $\nu$ at $\kappa=$ $0.080,\,$ $0.100,\,$ $0.120$ is larger than the value at tricritical point.

From the values of $\nu$ in Table.~\ref{critical_beta_B4},  we conclude that the nature of transition at $\kappa=0.060,\,$ $ 0.070$ is of first order. The values of $B_4(\beta_c,\infty)$ at $\kappa=0.060,\, 0.070$ are larger than   the expected  value.  This is because logarithmic scaling corrections will be present near the tricritical point~\cite{ls,deForcrand:2010he},  and our  simulations are carried out on finite size volume  on which large finite size corrections are observed  in simpler spin model~\cite{Billoire:1992ke}. However, the critical exponent $\nu$ is not sensitive to finite size corrections~\cite{deForcrand:2010he}.

We also present the reweighted distribution of the imaginary part of Polyakov loop at $\kappa=0.080,\,$ $0.120$ in
Fig.~\ref{fig6},  from which we can find that the behaviours of the reweighted distribution of ${\rm Im}(L)$  are in favor of second order transition. The behaviour of the reweighted distribution of ${\rm Im}(L)$ at $\kappa=0.100$ is also checked.

  Put our results with those in Ref.~\cite{Wu:2013bfa} together, we can estimate that the two tricritcal points are between $0.070<\kappa<0.080$ and $0.120<\kappa<0.140$, and when $\kappa<0.060$, our simulation results can not enable us to  determine the nature of transition decisively.

\section{DISCUSSIONS}\label{SectionDiscussion}

We have studied the nature of critical end points of Roberge-Weiss transition of two flavor lattice QCD with
Wilson fermions. When $i\mu_I = i\pi T$, the imaginary part of Polayakov loop is the order parameter for studying
the transition from low temperature phase to high temperature one.

In Ref.~\cite{Philipsen:2014rpa}, Wilson fermions are employed to study the nature of RW transition end points.
In Ref.~\cite{D'Elia:2009qz,Bonati:2010gi} and Ref.~\cite{deForcrand:2010he}, the simulations with staggered fermions show that  phase diagram of two flavor and three flavor QCD at imaginary
chemical potential $i\mu_I=i\pi T$ are characterized by two tricritical points, respectively.

 Our simulations are carried out at 8  values of $\kappa$ on $L_t=4 $ lattice on different 4 spatial volumes.   
 At each $\kappa$ value, we take the $\beta$ values in the reweighting procedure by monitoring the behavior of susceptibility of
 imaginary part of Polyakov loop.  As an example, we present in Fig.~\ref{fig7} the behavior of $\chi$ 
   and Binder cumulant $B_4$  with different selection of $\beta$ values on lattice $L_s=8$ at $\kappa=0.120$. From Fig.~\ref{fig7}, 
   we can find that the selection of $\beta=5.58,\,$ 
   $5.60,\,$ $5.62,\,$ $ 5.64,\,$ $5.66,\,$ $5.68$ may be the best. As a comparison, we present the results in the upper
   panels with $\beta=5.62,\,$ $5.64,\,$ $ 5.66,\,$ $ 5.68$, and in the lower panels with $\beta=5.58,\,$ 
  $ 5.60,\,$ $ 5.62,\,$ $ 5.64$. 
  
 At $\kappa=0.020,\,$ $0.040$, our simulations are not decisive for the determination of transition nature.

 At $\kappa=0.060,\,$ $0.070,\,$ $0.080,\,$ $0.100,\,$ $0.120,\,$ $0.140$, when  the behaviour of $\chi/{L_s^{\gamma/\nu}}$ is examined, it is difficult to decide the transition nature  at RW transition  end point. We turn to investigate Binder cumulant. By fitting Eq.~(\ref{binder_scaling_02}) to our data, we can extract the values of critical index $\nu$ which are collected in Table.~\ref{critical_beta_B4}, the $\nu$
 value at  $0.060,\,$ $0.070, \,$  $0.140\,$ supports that transition at the end points is of first order, whereas the $\nu$ value at
$0.080,\,$ $0.100, \,$  $0.120\,$ favors that the nature of transition is of second order.

By monitoring the change of $\nu$ at different $\kappa$ values and comparing these values with  those in the thermodynamic limit, we conclude that the tricritical points are between $0.070<\kappa<0.080$ and $0.120<\kappa<0.140$. However, this result is not in accord with the conclusion in Ref.~\cite{Philipsen:2014rpa} where Philipsen and Pinke believed that the two tricritical $\kappa$'s are $0.100(9)$ and $0.155(5)$.

Considering the lattice volume in our simulations at $\kappa=0.140$ is $8^3, \,$ $12^3$, our result of light tricritical point position is less reliable than that in  Ref.~\cite{Philipsen:2014rpa}. Moreover, it is interesting that  the position of heavy tricritical point and some $\nu,\,$ $\beta_{RW}$ values from our simulations are different from those in Ref.~\cite{Philipsen:2014rpa}, in spite of the same regularization we employed as that used by  Philipsen and Pinke.

We have no certain explanation for this disagreement, we speculate that it may be because of the different distribution  around $\beta_{RW}$
of those $\beta$ values
at which  simulations are carried out. In Ref.~\cite{Philipsen:2014rpa}, the $\beta$ values are more concentrated around $\beta_{RW}$ with $\Delta \beta=0.001$, whereas,  in our simulation, the $\beta$ values are more scattered around $\beta_{RW}$ with $\Delta \beta=0.02$.
Consequently, the $\beta$ values we used cover a wider range.   However, the less concentration  of  $\beta$ values may lead us to lose  much important information around  transition points, especially, when transition is under consideration.  With limited calculation resources, it may be better that more concentrated distribution of $\beta$ values around  transition points are used.

\begin{acknowledgments}
We thank Philippe de Forcrand for valuable helps.
 We modify the MILC collaboration's
public code~\cite{Milc} to simulate the theory at imaginary chemical
potential. In some of our calculation, we use the fortran-90 based multi-precision software~\cite{fortran}.   This work is supported by
the National Science Foundation of China (NSFC)  under Grant Nos.~(11105033, 11347029). The work was carried out at National Supercomputer Center in Tianjin, and the calculations were performed on TianHe-1(A).
\end{acknowledgments}

\end{document}